\documentclass[12pt,epsf]{article}
\usepackage{epsfig}

\newcommand{\setfigure}[2]{\begin{figure}[htbp]
\begin{center}\leavevmode\epsfxsize=5in\epsfbox{#1.eps}\end{center}\caption{#2\label{#1}}
\end{figure}}

\renewcommand{\thanks}[1]{\footnote{#1}} 

\newcommand{\be}{\begin{equation}}
\newcommand{\ee}{\end{equation}}
\newcommand{\bea}{\begin{eqnarray}}
\newcommand{\eea}{\end{eqnarray}}

\begin{document}

\pagestyle{empty}

\bigskip\bigskip
\begin{center}
{\bf \large Coordinates with Non-Singular Curvature for a Time Dependent
Black Hole Horizon}
\end{center}

\begin{center}
James Lindesay\footnote{e-mail address, jlslac@slac.stanford.edu} \\
Computational Physics Laboratory \\
Howard University,
Washington, D.C. 20059 
\end{center}
\bigskip

\begin{center}
{\bf Abstract}
\end{center}
A naive introduction of a dependency of the mass of a black
hole on the Schwarzschild time coordinate results in singular
behavior of curvature invariants at the horizon, violating
expectations from complementarity.  If instead a temporal
dependence is introduced in terms of a coordinate akin to
the river time representation, the Ricci scalar is nowhere
singular away from the origin.  It is found that for a
shrinking mass scale due to evaporation, the null
radial geodesics that generate the horizon are slightly
displaced from the coordinate singularity.  In addition,
a changing horizon scale significantly
alters the form of the coordinate singularity in diagonal
(orthogonal) metric coordinates representing the space-time.  A Penrose
diagram describing the growth and evaporation of an
example black hole is constructed to examine the evolution
of the coordinate singularity. 
\bigskip \bigskip \bigskip

\setcounter{equation}{0}
\section{Introduction}
\indent

There is a considerable interest\cite{Nielsen, Hajicek} in
understanding details of the
evaporation of black holes.  The singular
behavior of Schwarzschild coordinates
near the horizon makes those coordinates
inconvenient for describing the functional
dependencies of relevant physical parameters. 
One expects that a freely falling observer
should not encounter any particularly
singular behavior as the horizon is
traversed, since that observer's
coordinates manifest no singular
behavior at the Schwarzschild horizon. 
Therefore, it is useful to develop
coordinates that describe an evolution
without introducing invariant singular
behavior at the horizon.

Braunstein\cite{Braunstein} has examined the constraints
placed on the form of 
a conserved energy-momentum tensor in
a spherically symmetric geometry, and
shown that it must be singular at the
coordinate singularity associated with the
event horizon using his coordinates.  The author agrees
that there can be component singularities associated
with this tensor due to a particular coordinate
representation, but believes that
any true singularity associated with
physical scalars violates the principles of
complementarity and equivalence when
one compares the perspectives of
fiducial (fixed $r, \theta, \phi$) vs. freely
falling observers.  Generally, coordinates are constructed
to conveniently represent relationships
between events in particular frames of
reference, so that any singularities in scalar
physical parameters should be associated
with local physical content.  If there were
curvature singularities associated with
a slowly evaporating event horizon,
one would expect tidal effects producing a
type of ``brick wall", which is physically
unappealing.

Since the spatio-temporal behavior of 
Schwarzschild coordinates near
the Schwarzschild radius are particularly singular,
the anomalous singular behavior due to
those coordinates could be eliminated
by using a different time coordinate to
describe the shrinking mass scale. 
Such a time coordinate is introduced in
section \ref{nonsingularhorizon} based on the
non-orthogonal coordinates of the
river model for black holes\cite{rivermodel}. 
This time, which in a manner similar to that of
Schwarzschild corresponds to the time
of an asymptotic observer, has a behavior
near the horizon that does not introduce
new singular behavior in relevant scalar functional
forms.  Orthogonal coordinates will be developed
so that intuitive spatio-temporal relationships
can be associated with the evolution of a black
hole.  The behavior of coordinate measures near the coordinate singularity is
found to be significantly altered by any non-vanishing
temporal dependency $\dot{M} \not= 0$ in section \ref{diagsection}. 
A significant consequence of temporal dependency
is that the horizon is slightly displaced from the coordinate
singularity, as will be explored in section \ref{horizonsection}. 
To illustrate this displacement, a Penrose diagram for an example
black hole is constructed. 
In addition, holographic arguments are explored to estimate the
temperature associated with the evaporation process. 
Finally, the dynamics of a scalar field is examined, and a
temporally slowly varying solution is briefly explored.

\setcounter{equation}{0}
\section{Temporal Dependence of
Schwarzschild Mass Scale}
\indent

The Schwarzschild geometry is known to describe
a spherically symmetric static space-time.  It is
of interest to examine the geometry generated
by naively giving a dependency of the
Schwarzschild mass scale on the Schwarzschild
time parameter $t_S$
\be
R_M \equiv {2 G_N M(ct_S) \over c^2} =R_S .
\ee
The Schwarzschild radius is directly determined
by the mass scale, and for clarity will be labeled
by $R_M$.  For a metric form given by
\be  
ds^2 = -(1- {R_M(ct_S) \over r} )(dct_S)^2 + {dr^2 \over 1 - {R_M(ct_S) \over r} } +
r^2 (d \theta ^2 + sin^2 \theta \, d\phi ^2) 
\ee
the non-vanishing affine connections (with $x_S ^0 \equiv ct_S$)
are
\be
\begin{array}{c c c}
\Gamma_{0 0}^0 =  -{\dot{R}_M \over 2 (r - R_M)} , &
\Gamma_{0 r}^0 = {1 \over 2 r} \left ( {R_M \over r - R_M} \right ) , &
\Gamma_{r r}^0 ={r^2 \dot{R}_M \over 2 (r - R_M )^3}  , \\ \\
\Gamma_{0 0}^r = { R_M \left ( r - R_M  \right ) \over 2 r^3}   , & 
\Gamma_{0 r}^r = {\dot{R}_M \over 2 (r - R_M)} , & 
\Gamma_{r r}^r = -{1 \over 2 r} \left ( {R_M \over r - R_M} \right )  , \\ \\
\Gamma_{\theta \theta}^r = - \left ( r - R_M \right ) , &
\Gamma_{\phi \phi}^r = - \left ( r - R_M \right ) sin^2 \theta  , &
\Gamma_{r \theta} ^\theta = {1 \over r} , \\ \\
\Gamma_{\phi \phi} ^\theta = -cos \theta \, sin \theta , &
\Gamma_{r \phi} ^\phi = {1 \over r} , &
\Gamma_{\theta \phi} ^\phi = cot \theta .
\end{array}  
\ee
A comparison of geodesic motion from rest allows a
determination of the proper acceleration associated
with a fixed (fiducial) observer
\be
a_{proper} = {R_M c^2 \over 2 r^2}  \sqrt{{r \over r - R_M}},
\ee
which is seen to be singular at $r=R_M$.

The Ricci tensor takes the form
\be
(( \mathcal{R}_{\mu \nu} )) =
\left ( \begin{array}{c c c c}
{2 \dot{R}_M ^2 + (r-R_M) \ddot{R}_M \over
2 (r-R_M)^2 } & -{ \dot{R}_M \over r (r-R_M) } & 0 & 0 \\
-{\dot{R}_M \over r (r-R_M)} & -{r^2 (2 \dot{R}_M ^2 + (r-R_M) \ddot{R}_M) \over
2 (r-R_M)^4} & 0 & 0 \\
0 & 0 & 0 & 0 \\
0 & 0 & 0 & 0
\end{array} \right ) .
\ee
In these expressions, the dots represent derivatives with
respect to $ct_S$.
Several components of the Ricci tensor are seen to be singular
at the coordinate singularity $r=R_M$.  Of more significance,
the Ricci scalar
\be
\mathcal{R}= -{r (2 \dot{R}_M ^2 + (r-R_M) \ddot{R}_M) \over
2 (r-R_M)^3} 
\ee
is seen to be singular at the coordinate singularity for non-vanishing
$\dot{R}_M$.  The (mixed) Einstein tensor given by
\be
((\mathcal{G}_\mu ^\nu )) =
\left ( \begin{array}{c c c c}
0 & { \dot{R}_M \over  (r - R_M)^2 } & 0 & 0 \\
-{\dot{R}_M \over r^2} & 0 & 0 & 0 \\
0 & 0 & 
{ r ( 2 \dot{R}_M ^2 +(r - R_M) \ddot{R}_M ) \over 2( r - R_M ) ^3 }  & 0 \\
0 & 0 & 0 & { r ( 2 \dot{R}_M ^2 +(r - R_M) \ddot{R}_M ) \over 2( r - R_M ) ^3 }
\end{array} \right ) 
\ee
is likewise seen to manifest this singular behavior at $r=R_M$.

Einstein equations can be used to examine the behavior of any
energy-momentum densities:
\be
\mathcal{G}_{\mu \nu} = \mathcal{R}_{\mu \nu} -
{1 \over 2} g_{\mu \nu} \mathcal{R} = 
-{8 \pi G_N \over c^4} \mathcal{T}_{\mu \nu} .
\label{Einstein}
\ee
The curvature scalar is associated with the trace of the
energy-momentum tensor
\be
\mathcal{R} ={8 \pi G_N \over c^4} g^{\mu \nu} \mathcal{T}_{\mu \nu} .
\label{singularR}
\ee
The Ricci scalar $\mathcal{R}$ should be non-singular at $r=R_M$ if the
coordinate $R_M$ is only a coordinate anomaly.
Singularities introduced into components of
physical parameters due to coordinate transformations
$\mathcal{T}_{\hat{\mu} \hat{\nu}} = { \partial x^\alpha \over \partial \hat{x}^{\hat{\mu}} } 
\mathcal{T}_{\alpha \beta} { \partial x^\beta \over \partial \hat{x}^{\hat{\nu}} }$
have a different significance from those associated with
invariant physical content.  The singular behavior of the scalar
$\mathcal{R}$ at $r=R_M$ for non-vanishing $\dot{R}_M$
represents a singular structure associated with the local
space-time that must be reflected in the physical content
as shown in Eq. \ref{singularR}, and for this scalar function
is independent of the particular coordinate description.

\setcounter{equation}{0}
\section{A Singularity-free Horizon
\label{nonsingularhorizon}}

\subsection{The river model}
\indent

The so called ``river model" has been explored by
several authors\cite{rivermodel, FRWdS} to gain insight into
the dynamics of horizons.  The metric
takes an off-diagonal form generically given by
\be  
ds^2 = -(dct_R) ^2 + [dr - \beta (r) dct_R] ^2 +
r^2 (d \theta ^2 + sin^2 \theta \, d\phi ^2) .
\ee
The ``speed" $\beta$ has been interpreted by
some to be the speed of radial outflow of the
space-time ``river" through which objects move
using the rules of special relativity\cite{rivermodel}. 
The transformation 
\be
ct_R = ct_*  - \int ^ r {\beta (r') \over 1-\beta ^2 (r')}dr'
\ee
diagonalizes the metric, giving the form
\be  
ds^2 = -(1-\beta ^2 (r) )(dct_*) ^2 + {dr^2 \over 1 - \beta ^2 (r)} +
r^2 (d \theta ^2 + sin^2 \theta \, d\phi ^2) .
\ee
The river speed becomes luminal at the horizon associated
with the $(ct_* , r)$ coordinates.

For the present examination, the metric will take the form
\be
ds^2 = -\left (1-{R_M (ct_R) \over r} \right ) (dct_R) ^2 +
2 \sqrt{{R_M (ct_R) \over r}} dct_R \, \, dr + dr^2 + r^2 \, d \omega ^2
\label{metric}
\ee
where $d\omega^2 \equiv d\theta^2 + sin^2 \theta \, d\phi^2$. 
This space-time asymptotically corresponds with
Minkowski space similar to (but not necessarily identical to)
the behavior of a Schwarzschild geometry. 
The non-vanishing affine connections ($x^0 \equiv ct_R$)
are given by
\begin{center} $
\Gamma_{0 0}^r = {1 \over 2 r} \left [ \left ( {R_M \over r} \right ) -
\left ( {R_M \over r} \right ) ^{2} + \dot{R}_M \left ( {r \over R_M} \right ) ^{1/2} \right ] ,
$ \end{center}
\be
\begin{array}{c c c}
\Gamma_{0 0}^0 = {1 \over 2 r} \left ( {R_M \over r} \right ) ^{3/2} , &
\Gamma_{0 r}^0 = {1 \over 2 r} \left ( {R_M \over r} \right ) , &
\Gamma_{r r}^0 = {1 \over 2 r} \left ( {R_M \over r} \right ) ^{1/2} , \\ \\
 & \Gamma_{\theta \theta} ^0 = -r \left ( {R_M \over r} \right ) ^{1/2} , &
\Gamma_{\phi \phi} ^0 = -r \left ( {R_M \over r} \right ) ^{1/2} sin^2 \theta , \\ \\
\Gamma_{0 r}^r = -{1 \over 2 r} \left ( {R_M \over r} \right ) ^{3/2} , & 
\Gamma_{r r}^r = -{1 \over 2 r} \left ( {R_M \over r} \right )  , \\ \\
 & \Gamma_{\theta \theta}^r = - \left ( r - R_M \right ) , &
\Gamma_{\phi \phi}^r = - \left ( r - R_M \right ) sin^2 \theta  , \\ \\
\Gamma_{r \theta} ^\theta = {1 \over r} , &
\Gamma_{\phi \phi} ^\theta = -cos \theta \, sin \theta , \\ \\
 & \Gamma_{r \phi} ^\phi = {1 \over r} , &
\Gamma_{\theta \phi} ^\phi = cot \theta .
\end{array}  
\label{connections}
\ee
The mixed Einstein tensor
\be
((\mathcal{G}_\mu ^\nu )) =
\left ( \begin{array}{c c c c}
0 & 0 & 0 & 0 \\
-{\dot{R}_M \over r^2} & 
-{\dot{R}_M \over R_M r} \sqrt{R_M \over r} & 0 & 0 \\
0 & 0 & -{\dot{R}_M \over 4 R_M r} \sqrt{R_M \over r} & 0 \\
0 & 0 & 0 & -{\dot{R}_M \over 4 R_M r} \sqrt{R_M \over r}
\end{array} \right ) 
\ee
is seen to be non-singular away from $r=0$.
Likewise, the Ricci scalar
\be
\mathcal{R}={3 \dot{R}_M \over 2 r^2} \sqrt{r \over R_M}
\ee
is non-singular away from a physical singularity
at the origin. 
This represents the key point of appeal of this
approach.  The invariant curvature (and any
related invariant physical parameter) is
nowhere singular away from the origin.  Any
coordinate singularities manifest only in components
unique to that particular coordinate representation.

\subsection{Diagonalization of metric form for a dynamic mass scale
\label{diagsection}}
\indent

The construction of orthogonal temporal-radial coordinates
is appealing, especially with regards to one's intuitive
use of the coordinates.  If one attempts a coordinate transformation
of the form
\be
d ct_R = A(ct_R,r) \, d ct_D + \Delta(ct_R,r) \, dr \quad , \quad
dr_R = dr ,
\ee
the function $\Delta$ can be chosen to immediately diagonalize
the metric in Eq. \ref{metric} if it is of the form
\be
\Delta(ct_R, r) = { \sqrt{ {R_M (ct_R) \over r} } \over 1 - { R_M (ct_R) \over r}  }.
\label{Delta}
\ee
The diagonalized metric then takes the form
\be  
ds^2 = -(1- {R_M(ct_R) \over r} ) A^2 (ct_R,r) \, dct_D ^2 + 
{dr^2 \over 1 - {R_M(ct_R) \over r} } +
r^2 (d \theta ^2 + sin^2 \theta \, d\phi ^2) .
\label{diagonalmetric}
\ee

The transformed temporal coordinate $t_D$ must satisfy
\be
{\partial ct_D \over \partial ct_R} = {1 \over A(ct_R, r)} \quad , \quad
{\partial ct_D \over \partial r} = -{1 \over A(ct_R, r)} \left (
{ \sqrt{ {R_M (ct_R) \over r} } \over 1 - { R_M (ct_R) \over r}  } \right ) ,
\ee
along with the integrability condition
\be
{\partial \over \partial r} log A + \Delta {\partial \over \partial ct_R} log A =
{\partial \over \partial ct_R} \Delta .
\label{integrability}
\ee
If $\ddot{R}_M=0$, a solution can be demonstrated for the coordinate
transformation.  The reduced coordinate $\zeta$ will
be defined by $\zeta \equiv {R_M \over r}$. 
The coefficient $A$ is assumed to approach unity for vanishing $\dot{R}_M$,
giving the usual static Schwarzschild coordinates.  It is convenient to
assume a form for this coefficient
\be
log A(ct_R, r) = F(\zeta) \, \dot{R}_M .
\ee
The integrability condition Eq. \ref{integrability} gives the equation
\be
{\partial \over \partial \zeta} F(\zeta) =
{ {\partial \Delta \over \partial \zeta}  \over \dot{R}_M \Delta - \zeta }
\quad \textnormal{for} \quad  \ddot{R}_M=0.
\ee
Defining a finite coordinate $(ct_{Ro}, r_o)$,
the coefficient $A$ then satisfies
\bea
A(ct_R,r) = A(ct_R,r_o) \,exp  \int_{\zeta(ct_R, r_o)} ^{\zeta(ct_R, r)} \left [
{{\partial \Delta(\tilde{\zeta}) \over \partial \tilde{\zeta}} \, \dot{R}_M \, d \tilde{\zeta} \over
\Delta(\tilde{\zeta}) \, \dot{R}_M - \tilde{\zeta} }\right ]  ,
\nonumber \\
A(ct_R,r_o)= \,exp  \int_{\zeta(ct_{Ro}, r_o)} ^{\zeta(ct_R, r_o)} \left [
{{\partial \Delta(\tilde{\zeta}) \over \partial \tilde{\zeta}} \, \dot{R}_M \, d \tilde{\zeta} \over
\Delta(\tilde{\zeta}) \, \dot{R}_M - \tilde{\zeta} }\right ] ,
\eea
where the constant $A(ct_{Ro},r_o)$ is chosen to satisfy correspondence
between river and orthogonal coordinates independent of
the value of $\dot{R}_M$. 
Near the coordinate singularity $\zeta \rightarrow 1$, the function $\Delta(\zeta)$
from Eq. \ref{Delta} becomes singular and the temporal coefficient $A$
is seen to be likewise singular $A \rightarrow {1 \over 1-\zeta}$.  Examining the
metric form in Eq. \ref{diagonalmetric}, this suggests that a
changing radial mass scale $\dot{R}_M \not= 0 $ significantly
alters the form of the coordinate singularity in the diagonal coordinates. 
For instance, the transformation to a radial conformal coordinate
$d \Pi = {dr \over  \left ( 1-{R_M \over r}  \right )
A(ct_R ,r) }$ is seen to be non-singular at the
coordinate singularity in these coordinates unless $\dot{R}_M =0$
(in which case it becomes the radial tortoise coordinate).

\subsection{Radial proper distance and acceleration}
\indent

A measurement of radial proper distance involves the determination
of the distance between radially separated regions at simultaneous times
in orthogonal temporal-radial coordinates. In such
coordinates a direct interpretation can be given for
distant simultaneous measurements as well as
temporal measurements at a fixed spatial
position.  From calculations
in Eq. \ref{diagonalmetric}, this is seen to be given by
\be
d\rho = {dr \over \sqrt{1-{R_M \over r}}} .
\label{properdistance}
\ee
This formula applies external to the coordinate
singularity $r=R_M$.  
Geodesic radial motion from rest can be determined
using the connections previously calculated;
\be
{d^2 r \over dc \tau ^2} + {1 \over 2r} \left [
{R_M \over r} + \left ( {\dot{R}_M \over 1 - {R_M \over r}}
\right ) \left ( {r \over R_M}  \right )^{1/2} \right ] =0 ,
\label{rgeodesic}
\ee
\be
{d^2 c t_R \over dc \tau ^2} + {1 \over 2r} \left (
{R_M \over r} \right)^{3/2} \left ( { 1 \over 1 - {R_M \over r}}
\right )  =0 ,
\ee
which is useful for determination of the proper
acceleration near the coordinate singularity.
The proper acceleration
of a static fiducial observer located at
$(r,\theta,\phi)$ can be calculated using
Eqns. \ref{properdistance} and \ref{rgeodesic},
giving
\be
a_{proper}={R_M c^2 \over 2 r^2}
\sqrt{{r \over r - R_M}} \left ( 1 +
{r^2 \dot{R}_M \over (r-R_M) R_M}
\left ( r \over R_M \right ) ^{1/2}
\right ) .
\ee
For a slowly evaporating black hole, this form is seen
to imply a vanishing proper acceleration at a
value approximately given by $R_M / (1+\dot{R}_M)$.

\subsection{Evolution of the horizon and mass scale
\label{horizonsection}}
\indent

A black hole horizon is a light-like surface corresponding
to a finite area of  radially ``outgoing" null geodesics.  In
a static Schwarzschild geometry, these null geodesics
maintain a fixed finite radial coordinate away from the
physical singularity at $r=0$.  The general form for
null radial geodesics in the dynamic geometry specified in Eq. \ref{metric}
is given by
\be
{d r_\gamma \over dct_R} = - \sqrt{R_M \over r_\gamma} \pm 1.
\ee
Outgoing photons ($r_\gamma$ increases with $ct_R$)
traverse trajectories that correspond to the upper sign.
The radial coordinate of a \emph{dynamic} horizon must satisfy
\be
{d R_H \over dct_R} = - \sqrt{R_M \over R_H} + 1 \quad ,  \quad
R_H = {R_M \over \left (1 - \dot{R}_H  \right ) ^2} \quad .
\label{horizon}
\ee
This equation defines the temporal behavior
of the horizon as the radial mass scale $R_M$ varies.
Unlike the case for the static geometry, the horizon here is not given by
the radial mass scale $R_H \not= R_M$, since photons
instantaneously located at the radial mass scale $R_M$
are momentarily stationary in $r$, while the horizon
is not.  The horizon is the outermost set of null geodesics that
satisfy Eq. \ref{horizon} without reaching outgoing null infinity $\mathcal{I}^+$. 
Photons with $r_\gamma > R_H$, $\dot{r}_\gamma > \dot{R}_H$
will escape the singularity at $r=0$ (even if $r_\gamma < R_M$). 
This is the reason that care has been taken to
differentiate between the terms Schwarzschild radius, radial mass
scale, horizon, and coordinate singularity.

For completeness, the behavior of a radially
infalling spherical light-like shell is given by
\be
{d R_{sh} \over dct_R} = - \sqrt{R_M \over R_{sh}} - 1 \quad ,  \quad
R_{sh} = {R_M \over \left (1 + \dot{R}_{sh}  \right ) ^2} \quad .
\ee
These relations will be helpful in the construction of a model
black hole that forms at a finite time from energy collapse,
then evaporates away, as developed in section \ref{evolution}. 

\subsection{Holographic considerations}
\indent

Thermodynamic estimates can be made using
holographic arguments relating the horizon to
entropy.  The entropy associated with a horizon
of area $A$ will be assumed to satisfy
\be
S=k_B {A \over 4 G_N} {c^3 \over \hbar}=
k_B {\pi R_H ^2 \over G_N} {c^3 \over \hbar}.
\ee
Since the dominant form of the energy is assumed to
be due to the mass of the black hole, the energy $U$
will be taken to be
\be
U=M c^2 = {c^4 \over 2 G_N} R_M =
{c^4 \over 2 G_N} R_H \, \left (
1 - \dot{R}_H \right )^2.
\ee
For the present, it will be assumed that any
``pressure" contribution to the thermodynamics of
the geometry is negligible compared to the entropic
contribution.  In this case, the temperature
of the black hole is given by
\be
T={dU \over dS} = {\hbar c \over 4 \pi k_B} \left [
{(1-\dot{R}_H)^2 \over R_H} +
2 (1-\dot{R}_H) {d \dot{R}_H \over dR_H}
\right ] .
\label{temperature}
\ee
External to the region for which
$\dot{R}_M \sim 1-{R_M \over r}$, the geometry
behaves very similarly to a Schwarzschild
space-time.  Therefore, an estimation of the evaporation
rate will be made using the (not too) near $R_M$
behavior of fields in Schwarzschild geometry.  Massless particles near
the coordinate singularity encounter an effective
potential barrier\cite{blackholes}, with the
s-wave barrier height of the order $U_{barrier} \sim
{\hbar c \over R_M}$ and range $\sim {3 \over 2} R_M$.  The average energy of the
thermal quanta is of order $\sim k_B T$, with
an expected average rate of quanta over the
energy barrier of order $\sim U_{barrier}/ \hbar$. 
Therefore the rate of evaporation can be estimated by
$\dot{M} c^2 = {c^4 \over 2 G_N} \dot{R}_M \sim -{k_B T \over \hbar c} {U_{barrier}
\over \hbar} \sim -{\hbar c \over R_M ^2}$, which means that
$\dot{R}_M \sim - \left ( {L_P \over R_M} \right )^2 , \, |\dot{R}_M|<< 1$ (where $L_P$
is the Planck length $L_P ^2 \equiv \hbar G_N /c^3 = (\hbar/ M_P c)^2$).
This implies that ${d \dot{R}_H \over dR_H} \sim {1 \over R_M} 
\left ( {L_P \over R_M} \right )^2 << {1 \over R_M}$, which
means that the temperature given in Eq. \ref{temperature}
is dominated by the first term in the bracket.

To check the consistency of the assumption of the entropic
domination of thermodynamic energy,
in the first law of thermodynamics $dU=T\, dS - P \, dV$
the ``pressure" term can be estimated to be of order
$\sim \mathcal{T}_r ^r 4 \pi r^2 dr \sim \dot{M} c^2 \, dr \sim
{\hbar c \over R_M} {dr \over R_M} $, while the
entropy term is estimated using Eq. \ref{temperature}
to be $T \, dS \sim {\hbar c \over L_P} {dr \over L_P} $, which clearly
dominates the pressure term for horizons beyond the Planck scale.

\subsection{Evolution of a spherically symmetric neutral black hole
\label{evolution}}
\indent

It is of interest to examine the global structure of a growing
and evaporating black hole.  In a Penrose diagram, the space-time
structure is represented using conformal coordinates (with
light-like curves represented by lines with slope $\pm 1$)
and the entire space-time mapped onto a finite diagram.
It is not useful to construct a relevant separate finite Penrose diagram for a system
with constant $\dot{R}_M$, since all points are contained
within the coordinate singularity for either very late or very early times.  However
one should be able to patch together diagrams for a mass scale
that initially grows, then later evaporates away.  Consider
an infalling spherically symmetric photon shell that contains
an energy $M_o c^2$.  The region interior to this shell is essentially flat
by Birkoff's theorem for spherically symmetric geometries,
while the exterior region satisfies the
geometry associated with a spherically symmetric mass distribution
until that mass evaporates away. 
All elements of the infalling shell will eventually
cross the surface bounding the region
for which any outgoing light would eventually hit a
singularity at $r=0$. 
\setfigure{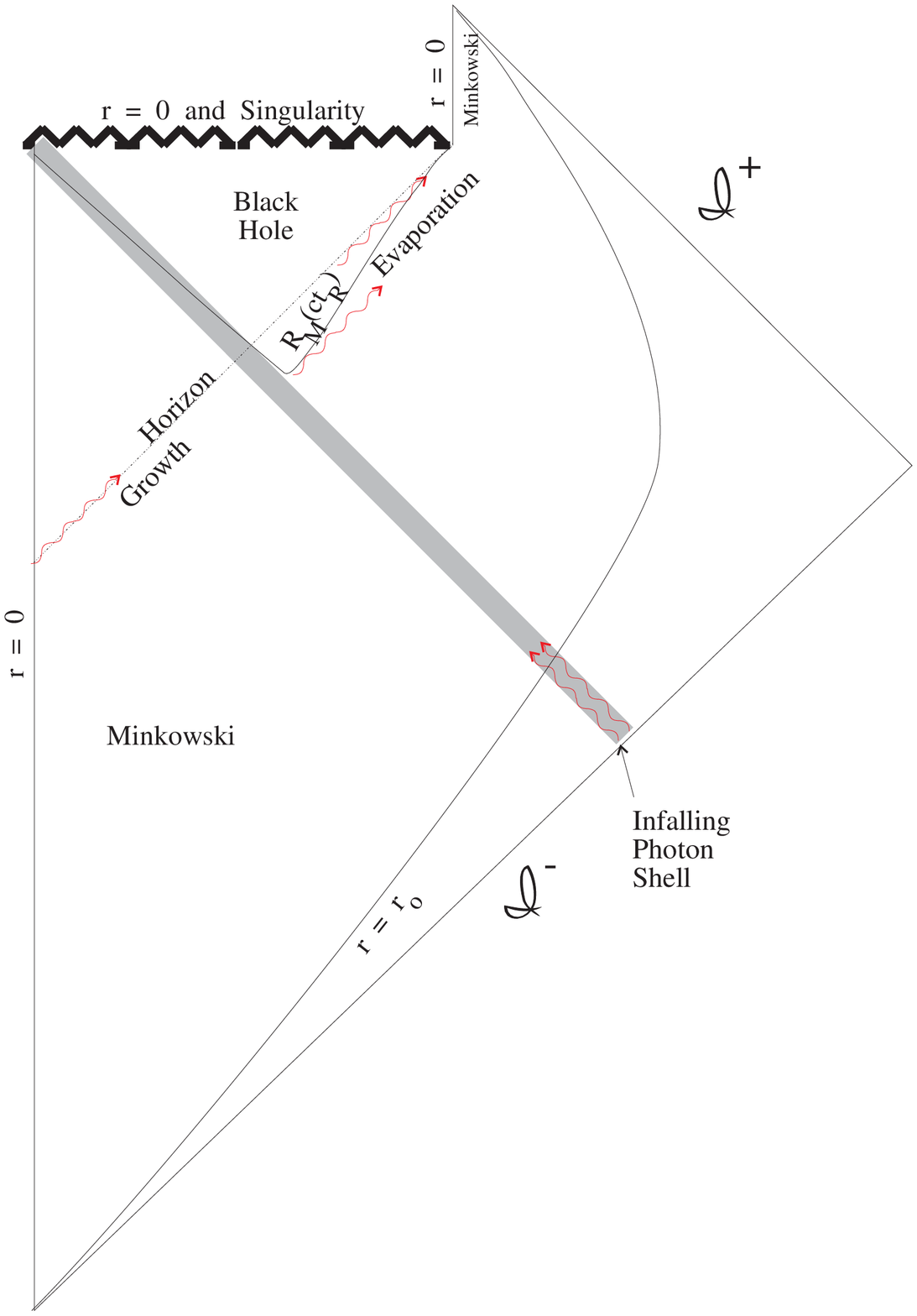}{Penrose diagram for formation and evaporation
of a spherically symmetric neutral black hole}

The Penrose diagram in Figure \ref{figure1} demonstrates
the expected global structure of such a space-time corresponding
to an initially flat (Minkowski) geometry with a radially
infalling photon shell of total energy $M_o c^2$.  The thick
band originating at $\mathcal{I}^-$ represents that
photon shell, and the region beneath that band (interior
to the shell) has
negligible curvatures due to Birkoff's theorem.  
This lower triangular
region is bounded on the left by the time-like curve
representing $r=0$.  Since the photon shell will eventually
reach $r=0$ forming a physical singularity, there will
be a light-like surface representing the innermost set of
out-going photons that can escape eventually hitting the
singularity which forms (indicated by the jagged horizontal
line on the diagram).  This light-like surface is
the horizon associated with the black hole, and it is
seen to be globally defined, having a
non-vanishing radial coordinate
$R_H > 0$ even prior to the space-time point(s) when
the infalling photon shell crosses this horizon.  However,
the radial mass scale $R_M$ associated with the
coordinate singularity in the highly curved metric
of the black hole geometry is seen to increase from a
vanishing value to that appropriate to a Schwarzschild-like
space-time as the photons in the shell cross this growing coordinate. 
As elements of the photon shell reach $r=0$, the curve $r=0 \, (< R_M)$
interior to the coordinate singularity becomes the
space-like singularity of increasing
strength represented by the initiation of the jagged curve. 
The width of this shell represents the duration
of the period of growth in the radial mass scale $R_M$. 
Increases in $R_M$ are associated with local
infalling shell photons as they cross growing
``horizon" scales (any of which would represent the
global horizon were the growth to stop at that stage),
and the curve $R_M(ct_R)$ grows
away from the physical singularity at $r=0$ after the
initial edge of the infalling photon shell initiates
this singularity.
In the space-time region for which the curved coordinates are of
relevance, the curve $r=0$ therefore tracks a physical singularity with
an associated mass scale $R_M \not= 0$.
The expected difference between the curve $R_M(ct_R)$ and
the horizon $R_H$ has been exaggerated for emphasis. 
This difference is determined by the relation for the
light-like curve given in Eq. \ref{horizon}.  
The curve $R_M(ct_R)$ crosses the global horizon
$R_H$ when $\dot{R}_H =0$, during which the rate of mass growth
is comparable to that of mass loss due to radiation. 
If the energy influx rate were to exactly match the evaporation
rate for an extended period, the geometry would be
expected to represent an essentially static Schwarzschild space.
For the case being examined, as
demonstrated in section \ref{horizonsection} a
photon emitted from $R_M$ is able to escape hitting
the singularity because of the shrinking of the mass scale
due to evaporation.  Since $R_M$ is associated
with the curved metric, radial coordinates relative to $R_M$ are determined
relative to the jagged singularity $r=0$.  During growth,
the coordinate singularity $R_M$ has a value less
than the radial coordinate of the horizon, whereas during evaporation
the horizon has radial coordinate less than the radial mass scale $R_H < R_M$. 
The physical singularity $r=0$ and the coordinate singularity
$R_M$ are seen to vanish together, leaving a (shifted)
time-like curve $r=0$ associated with the final low
curvature Minkowski-like space-time, represented as the
upper triangular region in the diagram subsequent
to complete evaporation of the singularity. 

From thermodynamic (first law) arguments, one
expects substantial modification from the entropic
dominated evaporation as $R_H$ approaches the
Planck scale.  During all periods with non-vanishing
radial mass scale, the singularity at $r=0$ \emph{is} a physical (space-like)
singularity hidden by the horizon $R_H$, which during
evaporation likewise lies within the coordinate singularity $R_M$. 
The physical singularity vanishes as $R_M \rightarrow 0$,
just as the horizon vanishes $R_H \rightarrow 0$.

\subsection{Scalar wave equation}
\indent

For completeness, the behavior of a free massless
scalar field in a geometry appropriately
parameterized by functional dependencies in
$(ct_R,r,\theta,\phi)$ will next be explored.  The
action will be assumed to take the form
\be
\mathcal{W}={1 \over 2} \int g^{\mu \nu}
\partial_\mu \chi^* \, \partial_\nu \chi \,
\sqrt{-g}  \, dx^0 dx^1 dx^2 dx^3 .
\ee
Substituting the decomposition
$\chi = \sum_{\ell m} {\psi_\ell (ct_R, r) \over r} Y_\ell ^m (\theta, \phi)$
results in a dynamical equation of the form
\be
\begin{array}{c}
-{\partial^2 \psi_\ell \over (\partial ct_R) ^2} \, \, + \, \,
{\partial \over \partial ct_R}
\left [ \sqrt{R_M \over r} \left ( 
{\partial \psi_\ell \over \partial r} \right ) \right ] \, \, + \, \,
{\partial \over \partial r}
\left [ \sqrt{R_M \over r} \left ( 
{\partial \psi_\ell \over \partial ct_R} \right ) \right ]
\, \, + \quad \quad \\ \\
{\partial \over \partial r}
\left [ \left ( 1-{R_M \over r} \right ) \left ( 
{\partial \psi_\ell \over \partial r} \right ) \right ]  -
\left [ {\ell (\ell + 1) + {R_M \over r}  \over r^2} +
{1 \over r} {\partial \over \partial ct_R} 
\sqrt{{R_M \over r}}\right ] \psi_\ell =0 .
\end{array}
\label{scalarwave}
\ee
In the special case of steady growth or evaporation
$\ddot{R}_M=0$, this equation admits a (temporally) slowly
varying solution in terms of the variable
$\zeta \equiv {R_M(ct_R) \over r}$ that will be briefly discussed. 
Substituting this variable dependency, Eq. \ref{scalarwave} takes
the form
\be
\begin{array}{c}
\left[ -\dot{R}_M ^2 - 2 \dot{R}_M \zeta^{3/2} +
(1-\zeta) \zeta^2 \right ] {\partial ^2 \psi_\ell \over \partial \zeta ^2} +
\left [ -3 \dot{R}_M \zeta^{1/2} + 2 \zeta - 3 \zeta^2 \right ]
{\partial \psi_\ell \over \partial \zeta} + \\ \\
- \left [ \ell (\ell + 1) + \zeta + {1 \over 2} \dot{R}_M \zeta^{-1/2} 
\right ] \psi_\ell =0.
\end{array}
\ee
This field has a near-horizon ($r \rightarrow R_M,
\, \zeta \rightarrow 1$) form given by $\psi_\ell ^{H} (\zeta) \approx 
\psi_\ell ^H (1) exp\left ( [ \ell (\ell + 1) +1](1-\zeta) \right )$, while for an
evaporating black hole, the asymptotic (s-wave) behavior $\zeta \rightarrow 0$
with vanishing field density is in
the form of a modified Bessel function
$\psi_0 ^\infty (\zeta) \sim \sqrt{\zeta} \, I_{{2 \over 3}}
({4 \over 3} {\zeta^{3/4} \over \sqrt{-2 \dot{R}_M}})$.  If such
a field were present in this geometry, its
amplitude would have to be of a scale
such that its energy density would have a negligible contribution
to the space-time geometry in order for this solution to be consistent.

\setcounter{equation}{0}
\section{Conclusions}
\indent

A temporal parameter $t_R$ that is not completely orthogonal to
the radial parameter has been shown to be adequate
for describing the evolution of the physical parameters
associated with a spherically symmetric black hole without
introducing invariant singular behavior away from
$r=0$. 
The coordinate singularity introduced by these asymptotically flat coordinates
is slightly displaced from the light-like surface that defines a finite
horizon which if crossed insures an eventual meeting with the
physical singularity at $r=0$.  During evaporation, this displacement occurs because
a radially outgoing photon originating from the coordinate
singularity is momentarily at fixed (then increasing) radial coordinate, while
the horizon has a slowly decreasing radial scale. 
This calculated discrepancy between the radial coordinates of
the horizon and the coordinate singularity, as well as the
modification in the form of the proper acceleration near
the horizon, give a natural basis for the scale of a
\textit{stretched horizon}, within which there is significant
modification of the local physics. 

Holographic considerations suggest that the thermodynamics of
evaporation is not significantly altered by the temporal dependence.
The behavior of a massless scalar field has been examined, and
besides the usual quantum modes, a particular solution
in terms of the parameter $\zeta = {R_M \over r}$ is
admitted in this geometry if $\ddot{R}_M=0$.
The radial coordinate measure of proper scale in these coordinates
behaves essentially the same as does that in Schwarzschild coordinates. 
However, there is a considerable difference in the behavior of the
orthogonal temporal coordinate measure $t_D$ to that of Schwarzschild coordinates
near the coordinate singularity. This qualitative discrepancy occurs for
any non-vanishing value of $\dot{R}_M$.  The transformation relationship
between the temporal parameters $t_R$ and $t_D$ demonstrate the
introduction of additional singular behavior into the orthogonal coordinate,
justifying the preferred use of $t_R$ to describe the physical dynamics.

\begin{center}
\textbf{Acknowledgment}
\end{center}

The author gratefully acknowledges Lenny Susskind
for teaching him about the subtleties of horizons and black holes. 
In addition, the author thanks J.D. Bjorken for
introducing him to the river model for black holes, and for
discussions involving the structure of horizons during
black hole formation.

\end{document}